\documentclass[final,nomarks]{dmtcs-episciences}

\usepackage[margin=2.5cm]{geometry}
\usepackage[utf8]{inputenc}

\usepackage{amsmath}
\usepackage{amsfonts}
\usepackage{amssymb}
\usepackage{amsthm}

\usepackage{booktabs}

\usepackage[citestyle=alphabetic,
            bibstyle=alphabetic,
            hyperref=true,
            url=false,
            isbn=false,
            backrefstyle=three,
            maxcitenames=3,
            maxbibnames=100,
            block=none,
            bibencoding=utf8,
            backend=biber]{biblatex}

\bibliography{distinct-vectors} %

\usepackage{hyperref}

\hypersetup{
  colorlinks=true,
  linkcolor=red!30!black,
  citecolor=green!30!black, 
  urlcolor=blue!30!black,
  bookmarksnumbered=true,
  pdflang={en}, breaklinks}

\usepackage[absolute]{textpos}  %

\usepackage{paralist}
\usepackage{scrextend}          %

\usepackage{graphicx}

\usepackage[sort&compress,nameinlink,noabbrev,capitalize]{cleveref}

\usepackage{needspace}

\usepackage{todonotes}
\usepackage[final,notref,notcite]{showkeys}

\newcommand{\bigO}{\ensuremath{\operatorname{O}}}%
\newcommand{\smallo}{\ensuremath{\operatorname{o}}}

\theoremstyle{plain}
\newtheorem{theorem}{Theorem}%

\newtheorem{conjecture}[theorem]{Conjecture}

\theoremstyle{definition}

\newcommand{\proofparagraph}[1]{\par\smallskip\emph{#1}}

\newcommand{\prob}[6]{%
  \needspace{3\baselineskip}
  \begin{quote}
    \begin{labeling}{#6}%
      \setlength\topsep{-.5ex} \setlength\itemsep{-.5ex}
    \item[#1]
    \item[\emph{#2}]#3
    \item[\emph{#4}]#5
    \end{labeling}%
  \end{quote}%
}

\newcommand{\probdef}[3]{\prob{#1}{Instance:}{#2}{Question:}{#3}{as}}

\newcommand{\poly}{\ensuremath{\operatorname{poly}}}

\newcommand{\C}{\ensuremath{\mathcal{C}}}

\newcommand{\pDV}{\textsc{Distinct Vectors}}

\title[A Double-Exponential Lower Bound for the Distinct~Vectors Problem]{A Double-Exponential Lower Bound\\for the Distinct~Vectors Problem\thanks{This research is part of a project that has received funding from the European Research Council (ERC) under the European Union's Horizon 2020 research and innovation programme (Grant Agreement 714704).}}

\author{Marcin Pilipczuk \and Manuel Sorge}
\affiliation{University of Warsaw, Poland}

\received{2020-02-05}
\revised{2020-07-06}
\accepted{2020-09-02}
\publicationdetails{22}{2020}{4}{7}{6072}

\keywords{feature selection, data mining, computational complexity, parameterized algorithms}

\begin{document}

\maketitle
\begin{abstract}
  In the (binary) \pDV\ problem we are given a binary matrix $A$ with pairwise different rows and want to select at most $k$ columns such that, restricting the matrix to these columns, all rows are still pairwise different. 
  A result by Froese et al. [JCSS] implies a $2^{2^{\bigO(k)}} \cdot \poly(|A|)$-time brute-force algorithm for \pDV.
  We show that this running time bound is essentially optimal by showing that there is a constant $c$ such that the existence of an algorithm solving \pDV\ with running time $2^{\bigO(2^{ck})} \cdot \poly(|A|)$ would contradict the Exponential Time Hypothesis. 
\end{abstract}

\begin{textblock}{20}(0, 13.5)
\includegraphics[width=40px]{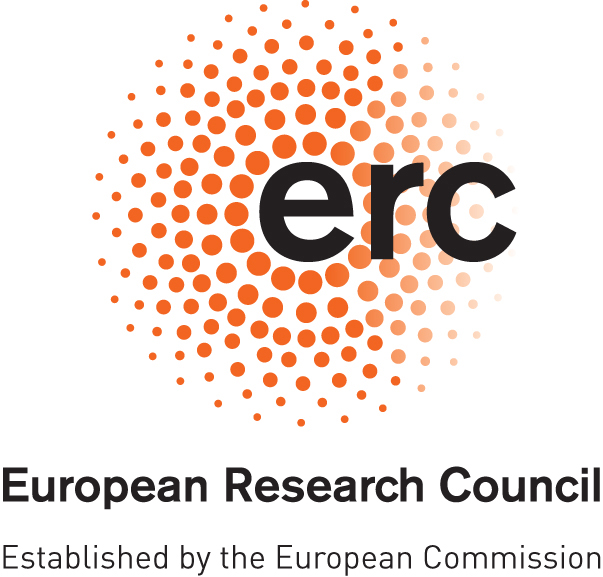}%
\end{textblock}
\begin{textblock}{20}(0, 14.5)
\includegraphics[width=40px]{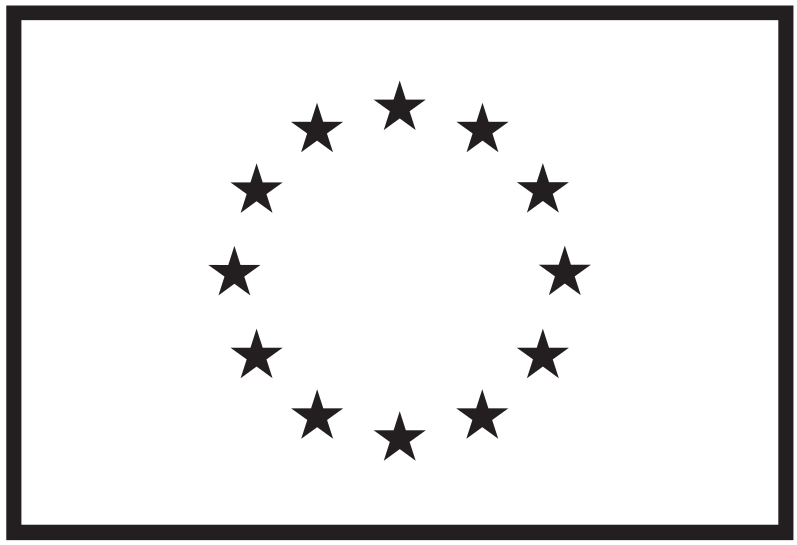}%
\end{textblock}

\section{Introduction}
For each $n \in \mathbb{N}$, let $[n] = \{1, \ldots, n\}$. Let
$\Sigma$ be a set and $n, m \in \mathbb{N}$. By $\Sigma^{m \times n}$
we denote the set of $m$-row $n$-column matrices with entries
in~$\Sigma$. Let $A \in \Sigma^{m \times n}$. By $A[i, j]$ we denote
the entry of $A$ in the $i$-th row and $j$-th column. By $A[i, *]$ and
$A[*, j]$ we denote the $i$-th row and the $j$-th column of $A$,
respectively. For easier notation, we often identify rows or columns
and their indices. Let $I \subseteq [m]$ and $J \subseteq [n]$. By
\begin{inparaenum}[(i)]
\item $A[I, J]$, 
\item $A[I, *]$, and
\item $A[*, J]$
\end{inparaenum}
we denote the submatrix of $A$ containing
\begin{inparaenum}[(i)]
\item only the entries that are simultaneously in rows in $I$ and
  columns in $J$,
\item only the entries in rows in $I$, and
\item only the entries in columns in $J$,
\end{inparaenum}
respectively.

We study the computational complexity of the following decision problem.

\probdef{\pDV}{A binary matrix $A \in \{0, 1\}^{m \times n}$ and
  $k \in \mathbb{N}$.}{Is there a subset $K \subseteq [n]$ of at most $k$
  columns such that the rows in $A[*, K]$ are pairwise distinct?}

We also say that $K$ as above is a \emph{solution}.

\pDV\ is a fundamental problem which has arisen in several different contexts.
Notably, it has applications in database theory, where it models key selection in relational databases (e.g. \cite{blasius_parameterized_2017}), machine learning, where it models combinatorial feature selection~\cite{charikar_combinatorial_2000}, and in rough set theory, where it models finding some minimal structure~\cite{pawlak_rough_1991}.
See \citeauthor{froese_finegrained_2018}~\cite{froese_finegrained_2018} for an overview over the literature.
We note that \pDV\ is sometimes formulated with larger alphabet size than two, that is, the entries of $A$ may be more than two distinct symbols.
Since we focus here on a lower bound, however, the binary formulation is sufficient for us.
\citeauthor{froese_exploiting_2016}~\cite[Theorem 12]{froese_exploiting_2016} gave a problem kernel with size $2^{2^{\bigO(k)}}$ for \pDV\ parameterized by~$k$.
(A problem kernel with respect to a parameter~$k$ is a polynomial-time self-reduction with an upper bound, a function of~$k$, on the resulting instance size.)  
Simple brute force on the resulting instances yields a $2^{2^{\bigO(k)}} \cdot \poly(|A|)$-time algorithm for \pDV.
It is natural to ask whether this running time bound can be improved. 
Here, we answer this question negatively by proving the following.

\begin{theorem}
  For each $\epsilon > 0$, if there is a $2^{\bigO(2^{ck})}\cdot \poly(n + m)$-time algorithm solving \pDV, then the Exponential Time Hypothesis is false, where $c = c(\epsilon) = 1/2 - \epsilon$.
\end{theorem}

Informally, the Exponential Time Hypothesis (ETH) states that \textsc{3SAT} on
$n$-variable formulas cannot be solved in $2^{o(n)}$~time~\cite{impagliazzo_complexity_2001}.
Formally, we rely on the following formulation that comes from 
an application of the Sparsification Lemma~\cite{impagliazzo_which_2001}.
\begin{conjecture}[Exponential Time Hypothesis + Sparsification Lemma]\label{c:ETH}
There exist constants $\delta, C > 0$ such that there is no algorithm
that, given as an input a 3CNF-SAT formula $\phi$ with $n$ variables and at most $C \cdot n$ clauses,
runs in time $\bigO(2^{\delta n})$ and correctly verifies the satisfiability of $\phi$.
\end{conjecture}

\noindent The proof of Theorem~1 is given in \cref{sec:lowerbound}. Herein, to simplify notation, we often write vectors
$(v_1, \ldots, v_n) \in \Sigma^n$ as $v_1v_2\ldots v_n$. We also use
$\cdot \circ \cdot$ to denote concatenation. That is, for each
$n, m \in \mathbb{N}$ and each $(v_i)_{i \in [n]} \in \Sigma^n$ and
$(w_i)_{i \in [m]} \in \Sigma^m$ we define
$v_1v_2\ldots v_n \circ w_1w_2\ldots w_m = v_1v_2\ldots
v_nw_1w_2\ldots w_m \in \Sigma^{n + m}$. Furthermore, for each
$i \in \mathbb{N}$ and $\sigma \in \Sigma$ we define
$\sigma^{(i)} = \sigma\sigma\ldots \sigma \in \Sigma^{i}$.
By $\log$ we refer to the base-two logarithm.
By $\poly$ we refer to an arbitrary fixed polynomial.

\section{Proof of Theorem 1}\label{sec:lowerbound}

\newcommand{\nvar}{\ensuremath{r}}
\newcommand{\ncla}{\ensuremath{s}}
  Let $\epsilon > 0$.
  Let $\delta$ and $C$ be the constants of Conjecture~\ref{c:ETH}.
  Let $\phi$ be a
  boolean formula $\phi$ in conjunctive normal form with
  $\nvar$~variables and $\ncla$~clauses such that each clause has size
  exactly three and such that $\ncla \leq C \cdot \nvar$.

  Below we construct an instance $(A, k)$ of \pDV\ which has a solution if and only if $\phi$ is satisfiable and such that $A$ has $n = 2^{\bigO(\nvar/\log\nvar)}$ columns and $m = \bigO(\nvar)$ rows, and there are $k \leq c' + 2\log\nvar$ columns to select for some constant~$c'$.
  The construction can be carried out in $2^{\bigO(\nvar/\log\nvar)}$~time.
  Thus, an algorithm solving \pDV\ with running time $2^{\bigO(2^{ck})}\cdot \poly(n + m)$ for some constant~$c$ can be used to check satisfiability of $\phi$ in time $2^{\bigO(\nvar/\log\nvar)} + 2^{\bigO(2^{c\cdot (c' + 2\log\nvar)})} \cdot \poly(2^{\bigO(\nvar/\log\nvar)} + \bigO(\nvar))$.
  Since $c = 1/2 - \epsilon$, this is $2^{o(r)}$ time, implying that the ETH is false.

  \proofparagraph{Construction.} %
  Let $X = \{x_1, x_2, \ldots, x_\nvar\}$ be the set of variables in
  $\phi$ and $\C = \{C_1, C_2, \ldots, C_\ncla\}$ the set of clauses.
  Without loss of generality, assume that $\nvar$ is a power of two
  and $\ncla$ equals $2^\ell - 1$ for some $\ell \in \mathbb{N}$.
  Otherwise, introduce variables that do not occur in any clause and
  repeat clauses as necessary. Note that this can be done in such a
  way that, afterwards, still $\ncla = \bigO(\nvar)$. Let
  $\nvar' := \lceil \nvar/\log\nvar \rceil$. We partition the
  variables into $\log \nvar$
  \emph{bundles}~$B_{i} = \{b_i^{1}, b_i^2, \ldots, b_i^{\nvar'}\}
  \subset X$, $i \in [\log\nvar]$, where each bundle~$B_i$ contains
  exactly $\nvar'$ variables (repeat variables from the bundle if necessary to fill a
  bundle).\footnote{We note that the construction works as long as the number of bundles is $\bigO(\log\nvar)$ and each bundle's size is~$\smallo(\nvar)$. We opted for $\log\nvar$ bundles as a natural choice.}

  The columns of matrix~$A$ are partitioned into $ \log(r) + 1$ parts,
  one \emph{consistency} part and one part for each bundle. The consistency
  part contains $\ell = \log(\ncla + 1)$ columns. We will make sure that
  all of them can be assumed to be in the solution. In this way, these
  columns will serve to distinguish some rows corresponding to clause
  gadgets from each other. The remaining $\log\nvar$ parts of columns
  correspond one-to-one to the bundles. The columns corresponding
  to $B_i$ are \emph{$B_i$'s columns}. For each $i \in [\log\nvar]$,
  there will be $\rho := 2^{\nvar'}$ columns belonging to $B_i$ which
  correspond one-to-one to the possible truth-assignments to the
  variables in~$B_i$. We will ensure that exactly one of the columns
  of~$B_i$ will be chosen in any solution, that is, the solution
  chooses a truth-assignment to the variables in~$B_i$.

  We now describe the construction of~$A$ by defining its rows. The
  rows of matrix $A$ are partitioned into two parts
  $I_1, I_2 \subseteq [m]$.
  \newcommand{\bin}{\ensuremath{\operatorname{bin}}}%

  Recall $\rho = 2^{\nvar'}$. The first part, $A[I_1, *]$, of the rows
  of~$A$ consists of $\log \nvar + 1$ rows, that is
  $I_1 = [\log \nvar + 1]$. The first row, $A[1, *]$, contains only
  zeros. The $(i + 1)$-th row, $i \in [\log \nvar]$,
  is defined by
  \[A[i + 1, *] = 0^{(\log(\ncla + 1))} \circ 0^{((i - 1)\rho)} \circ
    1^{(\rho)} \circ 0^{((\log\nvar - i)\rho)}.\] That is, for each
  bundle~$B_i$ there is a row which has 1 in the columns
  $\log(\ncla + 1) + (i - 1)\rho + 1$ to $\log(\ncla + 1) + i\rho$ and 0
  otherwise. We say that the columns $\log(\ncla + 1) + (i - 1)\rho + 1$ to
  $\log(\ncla + 1) + i\rho$ are the \emph{columns of bundle~$B_i$}. In
  order to distinguish the rows in~$I_1$ from the all-zero row, it is
  necessary, for each bundle~$B_i$, to pick at least one column in the
  set of columns belonging to~$B_i$ into the solution.

  \newcommand{\sat}{\ensuremath{\operatorname{sat}}} The second part,
  $A[I_2, *]$, of the rows of~$A$ consists of $2\ncla$ rows, that is
  $I_2 = \{\log\nvar + 2, \log\nvar + 3, \ldots, \log\nvar +
  2\ncla + 1\}$. For each $i, j \in \mathbb{N}$ with
  $1 \leq i \leq 2^j - 1$ let $\bin(i, j)$ be the binary $\{0, 1\}$-encoding of
  $i$ with exactly $j$ bits, padded with leading zeros if necessary.
  For each bundle~$B_i$, fix an ordering of the at most $\rho$~truth
  assignments to variables in~$B_i$.
  Recall that we may have repeated variables in~$B_i$.
  If so, then repeat truth assignments in the order fixed above so that their overall number is exactly~$\rho$.
  For each $p \in [\rho]$ and
  $q \in [\ncla]$, let $\sat_i(p, q) = 1$ if the $p$-th truth
  assignment makes clause~$C_q$ true and let $\sat_i(p, q) = 0$
  otherwise. Let $\sat_i(*, q) = (\sat_i(p, q))_{p \in [\rho]}$ and
  $\sat(q) = \sat_1(*, q) \circ \sat_2(*, q) \circ \ldots \circ
  \sat_{\log\nvar}(*, q)$. Define the $(2q - 1)$-th row in $I_2$,
  $q \in [\ncla]$, by
  \[ A[ \log\nvar + 2q, *] = \bin(q, \log(\ncla + 1)) \circ
    \sat(q).\]
  We call these rows the \emph{odd rows} in~$I_2$.
  Define the $2q$-th row in $I_2$, $q \in [\ncla]$, by 
  \[ A[ \log\nvar + 2q + 1, *] = \bin(q, \log(\ncla + 1)) \circ 0^{(n
      - \log(\ncla + 1))}.\] These are the \emph{even rows} in~$I_2$.
  We say that the $(2q - 1)$-th and the $2q$-th rows \emph{correspond}
  to clause~$q$.

  Finally, set $k = \log(\ncla + 1) + \log\nvar$. This concludes the
  construction of the \pDV\ instance~$(A, k)$.

  Before proving the correctness, observe that all our other
  requirements on the construction are satisfied: For the number~$k$ of
  columns to select, we have (recall that
  $\ncla \leq C \nvar$)
  \[k = \log(\ncla + 1) + \log\nvar \leq \log(2\ncla) + \log\nvar = \log(2C) + 2\log\nvar\text{.}\]
  Moreover, number $n$ of columns satisfies
  $n = \log(\ncla + 1) + \rho\log\nvar = 2^{\bigO(\nvar/\log\nvar)}$; and the
  number~$m$ of rows satisfies
  $m = 1 + \log\nvar + 2\ncla = \bigO(\nvar)$, each as required.
  Furthermore, since there are $2^{\bigO(\nvar / \log\nvar)}$ truth
  assignments to the variables in each bundle, the reduction can be
  carried out in $2^{\bigO(\nvar / \log\nvar)}$~time.

  \proofparagraph{Correctness.} We now prove that there is a solution
  to the above-constructed instance~$(A, k)$ of \pDV\ if and only if
  $\phi$ is satisfiable.

  Assume that $(A, k)$ has a solution~$K$.
  First, note that the even rows in $A[I_2, *]$ together with the all-zero row in~$I_1$ are $\ncla + 1$ rows that pairwise differ only in the first $\log(\ncla + 1)$ columns.
  Since for each $t \in \mathbb{N}$ we have that $t$ selected columns can pairwise distinguish at most $2^t$ rows, we thus have $[\log(\ncla + 1)] \subseteq K$.
  Let $K' = K \setminus [\log(\ncla + 1)]$ and observe $|K'| \leq \log\nvar$.
  Observe that in $A[I_1, *]$ there are $\log\nvar$ rows that each differ from the all-zero column in $A[I_1, *]$ only in the columns corresponding to some distinct bundle.
  Thus, for each bundle $B_i$, there is exactly one column, say $r_i$, in $K' \cap R_i$ where $R_i$ is the set of $B_i$'s columns, and no other columns are in~$K'$.
  Observe that each $r_i$ corresponds by construction to a truth assignment to variables in~$B_i$.
  Call this truth assignment~$\alpha_i$.
  Thus, taking the union over all $i \in [\log\nvar]$ of the truth assignment~$\alpha_i$ to the variables in $B_i$ represented by~$r_i$, we get a truth assignment~$\alpha$ to all variables in~$X$.
  This truth assignment~$\alpha$ is well-defined since the bundles constitute a partition of the variables.
  We claim that $\alpha$ satisfies $\phi$.

  Since $K$ is a solution, for each $q \in [\ncla]$, the sub-row
  $A[\log\nvar + 2q, K]$ is different from $A[\log\nvar + 2q + 1, K]$.
  These two sub-rows differ only in columns of bundles~$B_j$ that
  correspond to some truth assignment to the variables in~$B_j$ that
  satisfies clause~$C_q$. Thus, $\alpha$ satisfies $C_q$ and indeed,
  since this holds for all $q \in [\ncla]$, $\alpha$ satisfies $\phi$,
  as required.

  Now assume that there is a truth assignment~$\alpha$ to variables
  in~$X$ that satisfies~$\phi$. For each bundle~$B_i$, there is a
  column~$r_i$ in $B_i$'s columns such that the corresponding truth
  assignment, call it~$\alpha_i$, assigns to variables in~$B_i$ the
  same truth values as~$\alpha$. We construct a solution~$K$ to
  $(A, k)$ as follows. First, we put $[\log(\ncla + 1)] \subseteq K$. Then,
  for each bundle~$B_i$ put $r_i \in K$. This concludes the
  construction. Observe that $|K| = \log(\ncla + 1) + \log\nvar$, as
  required. It remains to show that all rows in $A[*, K]$ are
  distinct.

  Consider two rows $i, j \in [m]$, where $i \neq j$. We distinguish
  the following cases.
  \begin{asparaenum}[{Case} 1)]
  \item $i, j \in I_1$. Then, one of the two rows, say~$i$, has $1$ in
    the columns of some bundle and row~$j$ has $0$ in these columns.
    Since by construction $K$ contains exactly one column from the columns of each bundle, thus, $A[i, K] \neq A[j, K]$.

  \item $i \in I_1$ and $j \in I_2$. Observe that each row in $I_1$
    has only zeros in the first $\log(\ncla + 1)$ columns and each row
    in $I_2$ has at least one one in the first $\log(\ncla + 1)$
    columns. Thus, $A[i, K] \neq A[j, K]$.

  \item $i, j \in I_2$. If $A[i, K]$ and $A[j, K]$ differ in the first
    $\log(\ncla + 1)$ columns, then we are done. Otherwise, both $i$ and
    $j$ correspond to the same clause, say~$C_q$, and they are not
    both even or both odd rows. Say $i$ is an odd and $j$ is an even
    row. By the definition of~$K$, there is a bundle~$B_\ell$ and
    a column~$r_\ell$ such that $\alpha_\ell$ satisfies~$C_q$. Thus,
    $A[i, r_\ell] = 1 \neq 0 = A[j, r_\ell]$ by construction of the
    two rows.
  \end{asparaenum}
  Thus, $K$ is a solution to~$(A, k)$, as required.
  This concludes the proof.

  \section{Acknowledgments}
  We thank Vincent Froese and Irene Muzi for interesting and helpful discussions.
  We thank three anonymous reviewers for their insightful comments that improved the presentation of the paper.

\printbibliography

\end{document}